\documentclass[aps,prl,reprint,superscriptaddress]{revtex4-2}
\usepackage{graphicx}
\usepackage{dcolumn}
\usepackage{bm}
\usepackage{color}
\usepackage{amsmath, amssymb, amsfonts, mathrsfs}
\usepackage{physics}
\usepackage[colorlinks=true, linkcolor=blue, citecolor=blue, urlcolor=blue]{hyperref}
\usepackage{tikz}
\usepackage{pgfplots}
\usepackage{booktabs} 

\usetikzlibrary{arrows.meta, decorations.markings}

\begin{document}

\title{A Field-Theoretic Framework for Work Statistics and Universal Scaling in Non-equilibrium Phase Transitions}

\author{Yanbo Qiao}
\thanks{These authors contributed equally to this work.}
\affiliation{School of Physics, Peking University, Beijing 100871, China}

\author{Ruohan Xu}
\thanks{These authors contributed equally to this work.} 
\affiliation{School of Physics, Peking University, Beijing 100871, China}

\author{H. T. Quan}
\email{htquan@pku.edu.cn}
\affiliation{School of Physics, Peking University, Beijing 100871, China}
\affiliation{Collaborative Innovation Center of Quantum Matter, Beijing 100871, China}
\affiliation{Frontiers Science Center for Nano-optoelectronics, Peking University, Beijing, 100871, China}

\date{\today}

\begin{abstract}
We develop a field-theoretic framework for work statistics in $O(N)$ models driven through criticality. By analyzing the dynamic renormalization group flow of composite power operators, we find the Kibble-Zurek scaling laws as a natural consequence of the flow, and we derive the scaling of work cumulants relevant to Kibble-Zurek scaling of topological defects from first principles, bypassing heuristic freeze-out argument. This yields the universal scaling $c_n \sim \tau_Q^{-\alpha_n}$ for the $n$-th work cumulant density: isolated quantum systems exhibit a scaling of $\alpha_n = p(d+nz)\nu/(1+pz\nu)$, whereas open quantum and classical systems undergo a dimensional collapse to $\alpha_n = pd\nu/(1+pz\nu)$. Validated by exact Gaussian solutions and numerical simulations, our theory establishes a foundation for general work statistics far from equilibrium, thereby bridging stochastic thermodynamics and the renormalization group theory. 
\end{abstract}
\maketitle
\textit{Introduction.---}
The concept of universality in phase transitions stands as one of the most profound achievements of twentieth-century physics, asserting that diverse physical systems exhibit identical scaling behaviors near critical points, governed solely by spatial dimensionality and symmetry rather than microscopic details~\cite{vojtaQuantumPhaseTransitions2003,sachdevQuantumPhaseTransitions1999,vasilevFieldTheoreticRenormalization2004, tauberCriticalDynamicsField2014,cardyScalingRenormalizationStatistical1996}. Extending these principles from equilibrium to non-equilibrium processes remains a frontier of active research~\cite{hohenbergTheoryDynamicCritical1977,chouEquilibriumNonequilibriumFormalisms1985,zhongRenormalizationgroupTheoryFirstorder2017,nigroScalingPropertiesWork2019,siebererUniversalityDrivenOpen2025,mittalFermionQuantumCriticality2026,zhongFinitetimeScalingIts2011}. A seminal work in this context is the Kibble-Zurek mechanism (KZM), which provides a heuristic picture for understanding how a system driven through a continuous phase transition at a finite rate $v$ (proportional to the inverse driving duration $\tau_Q^{-1}$) falls out of equilibrium due to the divergence of the relaxation time~\cite{Kibble1976, Zurek1985,zurekCosmologicalExperimentsCondensed1996,zurekDynamicsQuantumPhase2005}. As the dynamics ``freeze out'', topological defects emerge with a density $\langle n_{\text{ex}}\rangle \sim \tau_Q^{-\alpha}$, where the exponent $\alpha = d\nu/(1+z\nu)$ is determined by the spatial dimension $d$ and the critical exponents $\nu$ and $z$~\cite{Zurek1985,zurekCosmologicalExperimentsCondensed1996,zurekDynamicsQuantumPhase2005,Adolfo2020PRR,campoUniversalityPhaseTransition2014,Polkovnikov2010APT,polkovnikovUniversalAdiabaticDynamics2005, dziarmagaDynamicsQuantumPhase2010,feiWorkStatisticsQuantum2020,delcampoUniversalStatisticsTopological2018,gomez-ruizFullCountingStatistics2020}. Although the spatial properties of such excitations have been extensively studied, these quantities are not always proper observables~\cite{polkovnikovColloquiumNonequilibriumDynamics2011, biroliKibbleZurekMechanismInfinitely2010}. On the contrary, some observables such as non-equilibrium work, excess energy as well as entropy production also encode essential information about the non-equilibrium process~\cite{polkovnikovColloquiumNonequilibriumDynamics2011}. Specifically, the statistics of the non-equilibrium work is currently attracting increasing attention~\cite{deffnerNonequilibriumWorkDistribution2008, talknerFluctuationTheoremsWork2007, silvaStatisticsWorkDone2008,funoPathIntegralApproach2018,feiWorkStatisticsQuantum2020,cavinaConvenientKeldyshContour2023,feiUniversalScalingWork2021,feiNonequilibriumGreensFunctions2020,hodsagiKibbleZurekMechanism2020}. Because work statistics capture the full counting statistics (FCS) of the universal fluctuations, they provide a fundamental lens for probing higher-order energy exchange near criticality.

Recent studies have explored quantum work statistics in several special cases. Solvable models~\cite{feiWorkStatisticsQuantum2020} suggest a universal scaling of work cumulants. This result is generalized to isolated quantum systems with conformal symmetry~\cite{ feiUniversalScalingWork2021}. However, the argument cannot be easily extended to open systems. In addition, it mainly relies on the ``adiabatic-impulse-adiabatic'' heuristic picture, postulating that an observable $\hat{O}$ scales with the Kibble-Zurek length as $\langle \hat{O} \rangle \sim \xi_\text{KZ}^{-\Delta_O}$~\cite{dasOldNewScaling2016,mondalNonequilibriumDynamicsQuantum2010,polkovnikovNonequilibriumDynamicsClosed2011,lamacraftPotentialInsightsNonequilibrium2011,gritsevUniversalDynamicsQuantum2010}, where $\Delta_O$ is the total scaling dimension of $\hat{O}$. While intuitively valid, this static ansatz bypasses the microscopic evolution and the renormalization group (RG) flow of the time-dependent driving protocol~\cite{biroliKibbleZurekMechanismInfinitely2010}. To establish a rigorous foundation for the above argument, Finite-Time Scaling (FTS) theories identify the driving rate $v$ as a relevant scaling field with a scaling dimension $y_v = z + 1/\nu$~\cite{zhongFinitetimeScalingIts2011,yinFinitetimeScalingDynamic2012,gongFinitetimeScalingLinear2010,yinNonequilibriumQuantumCriticality2014,zengFinitetimeScalingKibbleZurek2025}. Incorporating $v$ into the scaling equations demonstrates that power-law scaling is a natural consequence of the dynamic RG flow~\cite{zhongFinitetimeScalingIts2011}. Yet, a unified field-theoretic derivation extending this dynamic RG framework to quantum work statistics, rooted in a microscopic action, is still lacking. Because stochastic thermodynamics and RG theory have traditionally been two disparate fields, the thermodynamics of non-equilibrium phase transitions remains largely unexplored.

To compute non-equilibrium trajectory work, the characteristic function of work (CFW) has been formulated via path-integral~\cite{mallickFieldtheoreticApproachNonequilibrium2011, aronNonEquilibriumDynamics2018, yeoSymmetryItsBreaking2019,cavinaConvenientKeldyshContour2023, cavina2025unifying, cavina2026quantum} and perturbative non-equilibrium Green’s functions~\cite{feiNonequilibriumGreensFunctions2020, mohantaFullStatisticsNonequilibrium2023}. However, standard perturbative treatments are restricted to gapped systems and fail to capture the non-perturbative universal fluctuations that emerge in the far-from-equilibrium regime, especially in phase transition systems. Furthermore, recent adaptations employing symmetrized complex-time contours~\cite{aronNonEquilibriumDynamics2018, cavinaConvenientKeldyshContour2023} obscure the dynamic RG transformations required to derive universal scaling laws. 

To overcome these obstacles, we generalize the Keldysh formalism~\cite{kamenevFieldTheoryNonequilibrium2023, siebererKeldyshFieldTheory2016,altlandCondensedMatterField2010,langFieldTheoryDynamics2024,heylScalingUniversalityDynamical2015,Maciejko2007,marinoQuantumDynamicalField2016,Haehl2017JHEP} to provide a first-principle foundation for the FCS of work of the driven $O(N)$ model. By generalizing the path-integral approach for trajectory work~\cite{funoPathIntegralApproach2018}, we establish a unified framework to calculate work statistics in both isolated (quantum) and open (quantum and classical) systems. Applying the dynamic RG, we derive the universal scaling behaviors for the $n$-th order work cumulant densities, $c_n \sim \tau_Q^{-\alpha_n}$. Isolated quantum systems exhibit a scaling of $\alpha_n = p(d+nz)\nu/(1+pz\nu)$ driven by coherent multi-particle interferences, whereas open systems exhibit a scaling law of $\alpha_n = pd\nu/(1+pz\nu)$ which is independent of $n$. This demonstrates that the universal scaling behavior near criticality is a natural consequence of RG analysis rather than a phenomenological ansatz.

\textit{Formalism of Work Statistics.---}
To establish a unified framework for calculating work statistics, we consider a real $N$-component field $\boldsymbol{\phi}$ within a generic Caldeira-Leggett thermal environment. The total Hamiltonian is $\hat{H}_{\mathrm{tot}}(t) = \hat{H}_S(t) + \hat{H}_B + \hat{H}_{SB}$, where the bath and coupling terms vanish in the isolated limit. The non-equilibrium drive acts homogeneously and isotropically on the $O(N)$ system, governed by $\hat{H}_S(t) = \int d^d x \left[ \frac{1}{2} \boldsymbol{\Pi}^2 + \frac{1}{2}(\nabla\boldsymbol{\phi})^2 + \frac{1}{2}r(t)\boldsymbol{\phi}^2 + \frac{u}{4!}(\boldsymbol{\phi}^2)^2 \right]$ (with $\boldsymbol{\Pi}$ being the conjugate momentum). While our formalism accommodates arbitrary continuous driving protocol, our subsequent scaling analysis will focus on a generalized power-law driving ending at the critical point $r(t) = r_i [1 - (t/\tau_Q)]^p$ ($p > 0$), characterized by the driving rate $v_p \equiv r_i \tau_Q^{-p}$, where $r_i = r(0)$ is the initial mass.

Since $\partial_t \hat{H}_{\mathrm{tot}} = \partial_t \hat{H}_S$, the work done equals the net energy change of the universe. Under the two-point measurement (TPM) scheme, the CFW $\chi(\lambda)$ is obtained by tracing over the product of operators~\cite{funoPathIntegralApproach2018, yeoSymmetryItsBreaking2019, cavinaConvenientKeldyshContour2023,aronNonEquilibriumDynamics2018}:
\begin{equation}
    \chi(\lambda) = \Tr_{S+B} \left[ U_{\mathrm{tot}}e^{-i\lambda \hat{H}_{\mathrm{tot}}(0)} \rho_{\mathrm{tot}} U_{\mathrm{tot}}^\dagger e^{i\lambda \hat{H}_{\mathrm{tot}}(\tau_Q)} \right],
\end{equation}
where $\rho_\text{tot}$ is the initial normalized thermal state and  $U_{\mathrm{tot}}\equiv U_{\mathrm{tot}}(\tau_Q,0)$ is the unitary evolution. Consequently, the $n$-th order work cumulant density $c_n$ can be generated by the cumulant generating function  $w(\lambda) \equiv L^{-d} \ln \chi(\lambda) - i\lambda \Delta f$~\cite{feiWorkStatisticsQuantum2020}, such that $c_n \equiv \partial_{i\lambda}^n w |_{\lambda=0}$. Here, $L$ is the system size and $\Delta f$ is the equilibrium free energy density difference. 

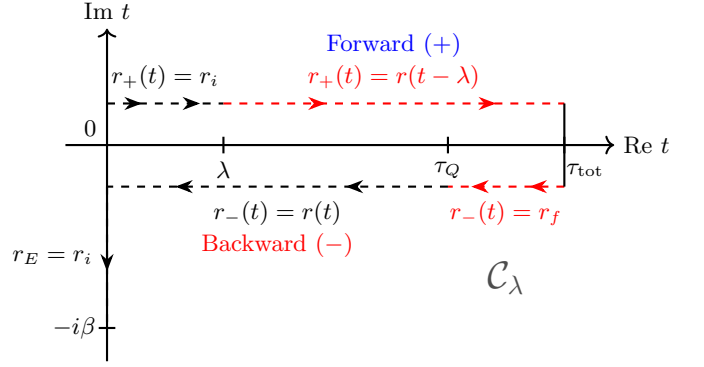
\begin{figure}[t]
    \centering
    \begin{tikzpicture}[scale=1.1, every node/.style={font=\small}]
        \draw[->, thick] (-0.5, 0) -- (6.1, 0) node[right] {$\text{Re } t$};
        \draw[->, thick] (0, -2.6) -- (0, 1.4) node[above] {$\text{Im } t$};
        \node at (-0.2, 0.2) {$0$};
    
        \tikzset{
            forward/.style={thick, dashed, postaction={decorate}, 
                decoration={markings, mark=at position 0.3 with {\arrow{Stealth}}, mark=at position 0.8 with {\arrow{Stealth}}}},
            backward/.style={thick, dashed, postaction={decorate}, 
                decoration={markings, mark=at position 0.3 with {\arrow{Stealth}}, mark=at position 0.8 with {\arrow{Stealth}}}},
            euclidean/.style={thick, dashed, postaction={decorate}, 
                decoration={markings, mark=at position 0.6 with {\arrow{Stealth}}}}
        }
    
        \draw[forward] (0, 0.5) -- (1.4, 0.5) node[midway, above=2pt] {$r_+(t) = r_i$};
        \draw[forward, red] (1.4, 0.5) -- (5.5, 0.5) node[midway, above=2pt] {$r_+(t) = r(t-\lambda)$};
        
        \draw[thick] (1.4, 0.1) -- (1.4, -0.1); 
    
        \draw[backward, red] (5.5, -0.5) -- (4.1, -0.5) node[midway, below=2pt] {$r_-(t) = r_f$};
        \draw[backward] (4.1, -0.5) -- (0, -0.5) node[midway, below=2pt] {$r_-(t) = r(t)$};
        
        \draw[thick] (4.1, 0.1) -- (4.1, -0.1);
    
        \draw[thick] (5.5, -0.5) -- (5.5, 0.5);
        \draw[thick] (5.5, 0.1) -- (5.5, -0.1);
    
        \node at (1.4, -0.3) {$\lambda$};
        \node at (4.1, -0.3) {$\tau_Q$};
        \node at (5.75, -0.3) {$\tau_{\text{tot}}$}; 
    
        \draw[euclidean] (0, -0.5) -- (0, -2.2) node[midway, left=2pt] {$r_E = r_i$};
        \draw[thick] (-0.1, -2.2) -- (0.1, -2.2) node[left=4pt] {$-i\beta$};
    
        \node[blue] at (3.45, 1.2) {Forward ($+$)};
        \node[red] at (2.05, -1.2) {Backward ($-$)};
    
        \node[font=\Large, text=black!70] at (4.8, -1.6) {$\mathcal{C}_\lambda$};

    \end{tikzpicture}
    \caption{Modified Schwinger-Keldysh contour $\mathcal{C}_\lambda$ and effective driving protocols for the FCS of work~\cite{funoPathIntegralApproach2018}. The real-time branches are extended to a total duration $\tau_{\text{tot}} = \tau_Q + \lambda$. The counting field $\lambda$ manifests as a relative temporal shift: the forward protocol $r_+(t)$ delays the driving protocol $r(t-\lambda)$ by maintaining the initial work parameter $r_i$ for $t \in [0, \lambda]$, whereas the backward protocol $r_-(t)$ follows the unshifted protocol and freezes at $r_f\equiv r(\tau_Q)$.}
\label{fig:contour_protocol}
\end{figure}

Evaluating this trace requires handling the non-commutative time-ordering $[\hat{H}_{\mathrm{tot}}(0), \hat{H}_{\mathrm{tot}}(\tau_Q)] \neq 0$. To deal with many-body systems, we generalize the real-time path-integral formalism introduced in~\cite{funoPathIntegralApproach2018}. The parameter $\lambda$—originating from TPM scheme~\cite{talknerFluctuationTheoremsWork2007} as the conjugate variable to trajectory work—emerges as a counting field, manifesting as a temporal shift that deforms the Schwinger-Keldysh contour $\mathcal{C}_\lambda$ along the real-axis (see Fig.~\ref{fig:contour_protocol}).

By formulating the dynamics on $\mathcal{C}_\lambda$, the global evolution is mapped onto an effective Keldysh generating functional for field $\boldsymbol{\phi}$ after integrating out the bath freedoms~\cite{kamenevFieldTheoryNonequilibrium2023, cavina2025unifying, cavina2026quantum}~\footnote{The boundary conditions enforce field continuity across the contour: the Euclidean branch dictates $\boldsymbol{\phi}_+(0, \mathbf{x}) = \boldsymbol{\phi}_E(\beta, \mathbf{x})$ and $\boldsymbol{\phi}_-(0, \mathbf{x}) = \boldsymbol{\phi}_E(0, \mathbf{x})$, while the quantum trace requires the endpoints to coincide, $\boldsymbol{\phi}_+(\tau_{\text{tot}}, \mathbf{x}) = \boldsymbol{\phi}_-(\tau_{\text{tot}}, \mathbf{x})$.}:
\begin{equation}
    \chi(\lambda) = \frac{1}{\mathcal{Z}(0)} \int_{\text{b.c.}} \!\!\! \mathcal{D}\boldsymbol{\phi} \exp \left( i S_+ - i S_- - S_E + i S_{\mathrm{diss}} \right).
\end{equation}
Here $\mathcal{Z}(0)$ is the normalization factor, $S_{\mathrm{diss}}$ encapsulates the bath-induced dissipation and thermal noise (naturally vanishes in isolated case) while $S_E$ represents the initial thermal state. The total duration $\tau_{\text{tot}} = \tau_Q + \lambda$ induces a temporal mismatch between the forward ($+$) and backward ($-$) branches:$$\begin{aligned} S_+ &= \int d^d x \left[ \int_0^{\lambda} \mathcal{L}_+ (r_i) dt + \int_{\lambda}^{\tau_{\text{tot}}} \mathcal{L}_+ (r(t-\lambda)) dt \right], \\ S_- &= \int d^d x \left[ \int_0^{\tau_Q} \mathcal{L}_- (r(t)) dt + \int_{\tau_Q}^{\tau_{\text{tot}}} \mathcal{L}_- (r_f) dt \right], \end{aligned}$$
where $\mathcal{L}_{\pm}$ denotes the system Lagrangian density $\mathcal{L}[\boldsymbol{\phi}_{\pm}, r]$. This deformed action encodes the field-theoretic FCS of work in both isolated and open systems.

\textit{Keldysh Rotation.---} To formulate the action, we apply the standard Keldysh rotation $\boldsymbol{\phi}_{\text{cl/q}} = (\boldsymbol{\phi}_+ \pm \boldsymbol{\phi}_-)/\sqrt{2}$, separating the classical field ($\boldsymbol{\phi}_{\text{cl}}$) from the quantum response field ($\boldsymbol{\phi}_{\text{q}}$). For the quartic term, this yields $\mathcal{L}_{\text{int}} \equiv -\frac{u}{12} [ \boldsymbol{\phi}_{\text{cl}}^2 (\boldsymbol{\phi}_{\text{cl}} \cdot \boldsymbol{\phi}_{\text{q}}) + \boldsymbol{\phi}_{\text{q}}^2 (\boldsymbol{\phi}_{\text{cl}} \cdot \boldsymbol{\phi}_{\text{q}}) ]$. The first term governs classical non-linear relaxation (recovering the Martin-Siggia-Rose-De Dominicis-Janssen (MSRDJ) formalism~\cite{langFieldTheoryDynamics2024,siebererKeldyshFieldTheory2016,kamenevFieldTheoryNonequilibrium2023}), while the second term captures quantum scattering, which leads to distinct universality class of the critical point.

\textit{Kinetic term and dynamic universality.---}The bare dynamics is governed by the kinetic operator $\mathcal{K} = \partial_t^2 + \gamma \partial_t - \nabla^2$~\cite{kamenevFieldTheoryNonequilibrium2023}. The inertial term ($\partial_t^2$) stems from unitary dynamics, while the dissipative friction ($\gamma \partial_t$) emerges from integrating out an Ohmic Caldeira-Leggett bath via the Feynman-Vernon functional~\cite{funoPathIntegralApproach2018, caldeiraIntroductionMacroscopicQuantum2014, smithGeneralizedFeynmanVernonApproach1987, breuerTheoryOpenQuantum2002, kamenevFieldTheoryNonequilibrium2023}. 
The dissipation strength $\gamma$ dictates the dynamic universality~\cite{langFieldTheoryDynamics2024}. In the isolated limit ($\gamma \to 0$), the inertial term dominates, preserving Lorentz invariance and yielding relativistic quantum critical dynamics ($z=1$). For any finite bath coupling ($\gamma > 0$), dissipation overwhelms inertial dynamics at late times ($t \gg \gamma^{-1}$). If we focus on the critical point and long-time physics, the dynamic RG flow will decouple: the open system is governed by dissipation, breaking Lorentz symmetry and transitioning to the overdamped Model A class ($z \approx 2$). This kinetic term distinguishes the isolated quantum and overdamped dissipative regimes~\cite{langFieldTheoryDynamics2024}.

\textit{Mass term and generating source.---}Work statistics are encoded within the mass term. The counting field $\lambda$ manifests as a temporal mismatch between the forward and backward protocols, $r_{\pm}(t)$. The mass Lagrangian density reads
\begin{equation}
    \mathcal{L}_{\text{mass}}(t) \equiv - 2\mathcal{M}_{\text{cq}} (\boldsymbol{\phi}_{\text{cl}} \cdot \boldsymbol{\phi}_{\text{q}}) - \mathcal{M}_{\text{cc}} \boldsymbol{\phi}_{\text{cl}}^2 - \mathcal{M}_{\text{qq}} \boldsymbol{\phi}_{\text{q}}^2,
    \label{eq:L_mass}
\end{equation}
where the classical-quantum mass is $\mathcal{M}_{\text{cq}}(t) \equiv [r_+(t) + r_-(t)]/4$, and the diagonal components are degenerate: $\mathcal{M}_{\text{cc}}(t) = \mathcal{M}_{\text{qq}}(t) \equiv [r_+(t) - r_-(t)]/4$. For $\lambda \to 0$, the paths coincide ($r_+ = r_-$) and $\mathcal{M}_{\text{cc}}, \mathcal{M}_\text{qq}$ vanish, recovering the standard Keldysh action where the CFW is normalized ($\chi=1$). Further, a counting field ($\lambda \neq 0$) breaks this normalization constraint, and these diagonal terms act as the generating source for the work statistics.

Expanding the mass term action with respect to $\lambda$ involves differentiating both the protocol and the integration limits~\cite{funoPathIntegralApproach2018}. By shifting the temporal dependence onto the fields~\footnote{See Supplemental Material~\cite{suppmat} for the derivation.}, the generating source simplifies to an operator over the driving period:
\begin{equation}
    S_{\mathrm{source}} = \sum_{n=1}^{\infty} \frac{\lambda^n}{n!} \int d^d x \int_0^{\tau_Q} dt \, \frac{1}{2} \dot{r}(t) \partial_t^{n-1} \left[ \boldsymbol{\phi}_+^2(t) \right],
\end{equation}
where the $n$-th order power operator is $\mathcal{P}^{(n)}(t) \equiv \frac{1}{2} \dot{r}(t) \partial_t^{n-1} [ \boldsymbol{\phi}_+^2(t) ]$. For the average work ($n=1$), $\mathcal{P}^{(1)} = \frac{1}{2}\dot{r}\boldsymbol{\phi}_+^2$ recovers the classical power operator. For $n \ge 2$, $\mathcal{P}^{(n)}$ incorporates temporal derivatives to capture quantum effects inherent to the TPM scheme.~\footnote{These higher-order insertions scale as $\mathcal{O}(\hbar)$ and vanish in the classical limit $\hbar \to 0$. See Supplemental Material~\cite{suppmat}.}

\textit{Adiabatic background.---}Although the source integrates over $t \in [0, \tau_Q]$, the temporal contributions far from the critical point ($r \gg 0$) yield a regular background evaluated via adiabatic perturbation theory (APT). In isolated systems, oscillatory phases suppress bulk transitions, localizing the non-adiabatic breakdown to the boundaries. This yields the scaling $c_1 \sim \tau_Q^{-2}$~\cite{polkovnikovUniversalAdiabaticDynamics2005, deffnerNonequilibriumWorkDistribution2008, dziarmagaDynamicsQuantumPhase2010, polkovnikovColloquiumNonequilibriumDynamics2011,Polkovnikov2010APT, feiWorkStatisticsQuantum2020} and $c_{n} \sim \tau_Q^{-2}$ for $n\geq 2$~\cite{feiWorkStatisticsQuantum2020, hodsagiKibbleZurekMechanism2020, solfanelliUniversalWorkStatistics2025, gaussian_manuscript_2026}. In open systems, connected correlation functions decay exponentially in the gapped phase~\cite{xuScalingBehaviorsWork2026}. This decay localizes the time coordinates within the $n$-point integrals, leaving the center-of-time as the extensive integration variable. The integration scales linearly with $\tau_Q$, yielding an APT scaling of $c_1 \sim \tau_Q^{-1}$~\cite{salamonMinimumEntropyProduction1980,xuScalingBehaviorsWork2026, mazonkaExactlySolvableModel1999, schmiedlEfficiencyMaximumPower2008, espositoEfficiencyMaximumPower2010, sivakThermodynamicMetricsOptimal2012} and $c_n \sim \tau_Q^{1-n}$ for $n \ge 2$~\cite{SheLeGeometricApproachNonequilibrium2018,xuScalingBehaviorsWork2026}.

\textit{Critical region.---}As the system approaches the phase transition, the energy gap vanishes, inducing the critical slowing down. The Keldysh action for this region combines the dynamic action with the generating source $S[\boldsymbol{\phi}_{\text{cl}}, \boldsymbol{\phi}_{\text{q}}, \lambda] \equiv S_+-S_-+S_\text{diss}$:
\begin{widetext}
\begin{equation}
    S[\boldsymbol{\phi}_{\text{cl}}, \boldsymbol{\phi}_{\text{q}}, \lambda] = \int d^d x \int dt \Big\{ -\boldsymbol{\phi}_{\text{q}} \big[ \mathcal{K} + r(t) \big] \boldsymbol{\phi}_{\text{cl}} + \mathcal{L}_{\text{diss}} + \mathcal{L}_{\text{int}} \Big\} + \sum_{n=1}^{\infty} \frac{\lambda^n}{n!} \int d^d x \int_0^{\tau_Q} dt \, \mathcal{P}^{(n)}(t).
    \label{eq:master_action}
\end{equation}
\end{widetext}
By the fluctuation-dissipation theorem, the Caldeira-Leggett bath induces a non-local Keldysh noise kernel~\cite{kamenevFieldTheoryNonequilibrium2023, cavina2026quantum}. The asymptotic limit of this noise determines the universality class. For isolated systems ($\gamma = 0$), the dissipation $\mathcal{L}_{\text{diss}}$ vanishes. For open systems at finite temperature, the infrared limit reduces the bath correlations to local Markovian white noise, $\mathcal{L}_{\text{diss}} \to i\gamma T \boldsymbol{\phi}_{\text{q}}^2$, yielding Model A dynamics. 

To extract the scaling behaviors of work cumulants, we analyze the scaling dimensions of the power operators $\mathcal{P}^{(n)}$ under the dynamic RG flow. We first consider the isolated quantum systems.

\textit{RG analysis for isolated systems.---}Setting the friction $\gamma \partial_t$ and the bath noise $\mathcal{L}_{\text{diss}}$ to zero, the bare kinetic operator preserves Lorentz invariance, enforcing dynamic exponent $z=1$. We denote the canonical dimension of a generic observable $\mathcal{X}$ as $[\mathcal{X}]$. Choosing canonical dimensions $[x] = -1$ and $[t] = -z$, the Keldysh fields scale symmetrically as $[\boldsymbol{\phi}_{\text{cl}}] = [\boldsymbol{\phi}_{\text{q}}] = (d+z-2)/2$ (hereafter simply denoted as $\boldsymbol{\phi}$). The coupling constant has a canonical dimension of $[u] = 4 - (d+z) \equiv \epsilon$. This identifies the upper critical dimension as $D_c = 4$, allowing us to perform RG analysis via the $\epsilon$-expansion.

Near the critical point, shifting the time coordinate to $s = \tau_Q - t$ yields the protocol $r(s) = v_p s^p$, where the driving rate $v_p \equiv r_i \tau_Q^{-p}$ possesses the canonical dimension $[v_p] = pz + 2$.

To capture critical fluctuations near the RG flow fixed point $u^*$, defined by the zero of the beta function $\beta_u \equiv \frac{du}{d\ln{\mu}} = 0$ (with $\mu$ being an arbitrary momentum scale), we relate bare quantities ($\mathcal{X}_0$) to the renormalized ones ($\mathcal{X}_R$) via the multiplication of renormalization factors, $\mathcal{X}_0 = Z_{\mathcal{X}} \mathcal{X}_R$. The corresponding anomalous dimension is then evaluated as $\gamma_{\mathcal{X}}^* = \mu \partial_\mu \ln Z_{\mathcal{X}} |_{u^*}$. The total scaling dimension is given by $\Delta_{\mathcal{X}} = [\mathcal{X}] + \gamma_{\mathcal{X}}^*$.

We determine the renormalization of the driving rate via its UV structure. Because the driving protocol $r(s) = v_p s^p$ is spatially uniform, it introduces no novel UV divergences beyond the static $O(N)$ theory. Consequently, the local operator $r(s)\boldsymbol{\phi}^2$ shares the same counterterm and the renormalization constant as the static mass insertion $r_i\boldsymbol{\phi}^2$. This enforces their anomalous dimensions to $\gamma_{v_p}^* = \gamma_r^* = 1/\nu - 2$. This yields the total driving rate dimension $\Delta_{v_p} = pz + 1/\nu$.

Similarly, the invariance of the action requires the bare mass insertion to match the renormalized one, $r_0 (\boldsymbol{\phi}^2)_0 = r (\boldsymbol{\phi}^2)_R$. It enforces $Z_{\boldsymbol{\phi}^2} = Z_r^{-1}$, which leads to $\gamma_{\boldsymbol{\phi}^2}^* = -\gamma_r^*$. The total scaling dimension of the composite operator $\boldsymbol{\phi}^2$is thus $\Delta_{\boldsymbol{\phi}^2} = d + z - 1/\nu$.

To map microscopic fluctuations to macroscopic observables, we combine the Callan-Symanzik equation with Euler’s homogeneous function theorem. At the Wilson-Fisher fixed point ($u \to u^*$, $\beta_u \to 0$), the beta function of driving rate is linearized as $\beta_{v_p} \to -\Delta_{v_p} v_p$, and the dynamic RG equation for the $n$-point connected correlator $G_R^{(n)}$ of an arbitrary operator $\mathcal{O}$ is given by:
\begin{equation}
\left[ \sum_{j=1}^n \left( \mathbf{x}_j \cdot \nabla_{j} + z t_j \partial_{t_j} \right) - \Delta_{v_p} v_p \frac{\partial}{\partial v_p} + n \Delta_{\mathcal{O}} \right] G_R^{(n)} = 0,
\end{equation}
where $j$ indexes the spacetime coordinates $(\mathbf{x}_j, t_j)$ of the $j$-th operator insertion. Integrating this equation reveals that $G_R^{(n)}$ is a generalized homogeneous function. Specifically, for the equal-time spatial two-point correlator $G_R^{(2)}(\mathbf{x})$, the flow yields the scaling form $G_R^{(2)}(\mathbf{x}; v_p) = b^{-2\Delta_{\phi}} \mathcal{G}(\mathbf{x}/b, v_p b^{\Delta_{v_p}})$\footnote{Integrating the spatial flow via $\hat{D} = -|\mathbf{x}|\partial_{|\mathbf{x}|} - 2\Delta_\phi$ rescales a short-range correlator $\sim e^{-|\mathbf{x}|/r_0}$ into $b^{-2\Delta_\phi} \mathcal{G}(|\mathbf{x}|/b) e^{-|\mathbf{x}|/(r_0 b)}$.}~\cite{zhongFinitetimeScalingIts2011}.
Setting $b = \xi_{v_p} \equiv v_p^{-1/\Delta_{v_p}}$ which absorbs the driving-rate dependence, we identify the correlation length and the relaxation time:
\begin{equation}
\xi_{v_p} \equiv v_p^{-\frac{1}{\Delta_{v_p}}}\sim \tau_Q^{\frac{p\nu}{1+pz\nu}}, \quad \tau_{v_p} \sim \xi_{v_p}^z.
\label{eq:KZM}
\end{equation}
Therefore, the $n$-point correlator for any operator $\mathcal{O}$ scales as $G_R^{(n)} \sim \xi_{v_p}^{-n\Delta_{\mathcal{O}}} \mathcal{G}^{(n)}(\{ \mathbf{x}_j \xi_{v_p}^{-1}, t_j \tau_{v_p}^{-1} \}; 1)$. This reveals the Kibble-Zurek scaling laws as a natural consequence of the RG flow.

Having established the critical scaling of correlators, we evaluate the generating functional of non-equilibrium work. The power operators $\mathcal{P}^{(n)}$ are defined based on the composite field $\boldsymbol{\phi}_+^2$.

Because the action is dimensionless, the scaling dimension of the counting field is fixed by the first term, $\lambda \int dt d^d x \, \mathcal{P}^{(1)}$.  Since $\Delta_{\mathcal{P}^{(1)}} = d + 2z$, we deduce that $\Delta_\lambda = -z$. 

For $n \ge 2$, the operators $\mathcal{P}^{(n)} \propto \partial_t^{n-1} \boldsymbol{\phi}_+^2$ possess scaling dimensions $\Delta_{\mathcal{P}^{(n)}} = d + (n+1)z$. Paired with the counting field $\lambda^n$, the combined insertion $\lambda^n \int dt d^d x \, \mathcal{P}^{(n)}$ yields zero net scaling dimension. They contribute to the higher-order work cumulants in the slow-driving limit, alongside the connected correlations of $\mathcal{P}^{(1)}$.

The work statistics are captured by the cumulant generating function density $w$. Given its canonical dimension $[w] = d$, rescaling the system by the correlation length $\xi_{v_p}$ enforces the universal scaling form $w = \xi_{v_p}^{-d} \Phi_w (\lambda \xi_{v_p}^{-z})$. Extracting $c_n = \partial_{i\lambda}^n w |_{\lambda=0}$ pulls down a scaling factor $\xi_{v_p}^{-z}$ for each derivative:
\begin{equation}
    c_n = \xi_{v_p}^{-d} \left( \xi_{v_p}^{-z} \right)^n \Phi_w^{(n)}(0) \sim \xi_{v_p}^{-(d+nz)}.
\end{equation}
The $n$-th derivative generates connected correlations of the operators $\mathcal{P}^{(k)}$. Here any operator combination satisfying $\sum k_i = n$ shares the identical scaling factor $\xi_{v_p}^{-nz}$. The amplitude $\Phi_w^{(n)}(0)$ absorbs the integration of these correlations, leaving the exponent unaffected.

Substituting Eq.~\eqref{eq:KZM} into $c_n$ yields the scaling law $c_n \sim \tau_Q^{-\alpha_n(p)}$:
\begin{equation}
    \alpha_n(p) = \frac{p(d + nz)\nu}{1 + pz\nu}.
    \label{eq:final_alpha_n}
\end{equation}
Equation~\eqref{eq:final_alpha_n} generalizes the KZM to the FCS of non-equilibrium work. This universal exponent recovers exact solutions of the 1D quantum Ising model~\cite{feiWorkStatisticsQuantum2020}, conformal field theory predictions~\cite{feiUniversalScalingWork2021}, and numerical simulations~\cite{hodsagiKibbleZurekMechanism2020}. One can also verify this result in the free field theory, see Ref.~\cite{gaussian_manuscript_2026}.

\textit{RG analysis for open systems.---}
To complete our scaling theory, we analyze the open dissipative systems via our formalism. While the general RG framework mirrors the isolated case, environmental coupling at finite temperature ($T>0$) introduces deviations that alter the scaling behavior. 

As established, the critical dynamics of the open system are governed by the friction $\gamma \partial_t$ and Markovian noise $i \gamma T \boldsymbol{\phi}_{\text{q}}^2$. Balancing these effective terms breaks the Lorentz symmetry of the Keldysh fields, enforcing $[\boldsymbol{\phi}_{\text{q}}] = [\boldsymbol{\phi}_{\text{cl}}] + z$~\cite{adzhemyanMultiloopCalculationsParametric2025}. Because the classical field retains a lower canonical dimension,  $[\boldsymbol{\phi}_{\text{cl}}] = (d-2)/2$, quantum insertions including $\boldsymbol{\phi}_{\text{q}}^2$ become irrelevant under the dynamic RG flow. The generalized power operators are thus governed by the classical composite field $\boldsymbol{\phi}_{\text{cl}}^2$.

To extract work statistics, the classical work term ($n=1$) in the action, $\int dt d^d x \lambda \dot{r}(t) \boldsymbol{\phi}_{\text{cl}}^2$, must be dimensionless, which leads to $\Delta_\lambda = 0$. This is because at the finite-temperature fixed point, the thermal bath sets the energy scale, rendering the counting field dimensionless.

For $n \ge 2$, each temporal derivative increases the scaling dimension of $\mathcal{P}^{(n)}$ by $z$. Because $\lambda$ has zero scaling dimension, the combined insertion $\lambda^n \int dt d^d x \, \mathcal{P}^{(n)}$ acquires a positive scaling dimension of $(n-1)z$. These terms are irrelevant under the dynamic RG flow. In the slow-driving limit, these terms vanish, scaling as $\xi_{v_p}^{-(n-1)z} \to 0$. The cumulant generating function density is determined solely by the $n=1$ term, scaling as $w = \xi_{v_p}^{-d} \Phi_w(\lambda)$. Extracting $c_n = \partial_{i\lambda}^n w |_{\lambda=0}$ yields no additional scaling factors, fixing the $n$-independent exponent:
\begin{equation}
    \alpha_n(p) = \frac{pd\nu}{1+pz\nu}.
   \label{eq:scaling_open}
\end{equation}
As far as we know, this result Eq.~\eqref{eq:scaling_open} has not been reported previously. This result agrees with numerical simulations of classical Ising models~\cite{suppmat}.

\textit{Competition between KZM and APT.---}The measured work cumulants accumulate both universal KZM singularities and regular APT contributions from gapped regions:
\begin{equation}
    c_n^{\text{total}} \approx \mathcal{A}_n \tau_Q^{-\alpha_n(p)} + \mathcal{B}_n \tau_Q^{-\beta_n},
\label{eq:scaling_competition}
\end{equation}
where $\mathcal{A}_n$ and $\mathcal{B}_n$ are non-universal amplitudes. The background exponent $\beta_n$ is governed by UV physics: $\beta_n = 2$ for isolated unitary dynamics~\footnote{The $\beta_n=2$ scaling assumes an abrupt initiation or termination of the drive ($\dot{r}(0) \lor \dot{r}(\tau_Q) \neq 0$ at least for one of the boundaries). For smoothed protocols where boundary derivatives vanish, this regular background is further suppressed, rendering the universal KZM scaling more dominant in the observable statistics.}~\cite{polkovnikovUniversalAdiabaticDynamics2005, deffnerNonequilibriumWorkDistribution2008, dziarmagaDynamicsQuantumPhase2010, polkovnikovColloquiumNonequilibriumDynamics2011,Polkovnikov2010APT, feiWorkStatisticsQuantum2020, hodsagiKibbleZurekMechanism2020, solfanelliUniversalWorkStatistics2025, gaussian_manuscript_2026}, whereas open dissipative friction yields $\beta_1 = 1$~\cite{salamonMinimumEntropyProduction1980,xuScalingBehaviorsWork2026, mazonkaExactlySolvableModel1999, schmiedlEfficiencyMaximumPower2008, espositoEfficiencyMaximumPower2010, sivakThermodynamicMetricsOptimal2012} and $\beta_{n} = n - 1$ ($n\geq 2$)~\cite{SheLeGeometricApproachNonequilibrium2018, xuScalingBehaviorsWork2026}. In the slow-driving limit ($\tau_Q \to \infty$), the KZM dominates only if $\alpha_n(p) < \beta_n$; otherwise, it is overwhelmed by the APT background (or acquires a logarithmic correction $c_n \sim \tau_Q^{-\alpha_n} \ln \tau_Q$ at $\alpha_n = \beta_n$~\cite{feiWorkStatisticsQuantum2020}). Furthermore, the p-dependence of $\alpha_n(p)$ allows one to engineer non-linear drives (by tuning $p$) to enforce the KZM-dominated regime, ensuring the universal statistics remain experimentally observable.

\textit{Summary.---}
We have developed a field-theoretic framework for the FCS of non-equilibrium work. By identifying the counting field $\lambda$ as a temporal shift on the Keldysh contour, we systematically map the TPM work onto dynamic RG flows. This establishes a theoretical bridge between stochastic thermodynamics and RG theory. Applying this framework, we derive the universal scaling for work cumulants. In isolated quantum systems, the scaling exponent relevant to KZM is $\alpha_n = p(d+nz)\nu/(1+pz\nu)$. In open systems, it results in an $n$-independent exponent $\alpha_n = pd\nu/(1+pz\nu)$. The total scaling behavior of work cumulants is the combination of $\alpha_n$ and $\beta_n$ (Eq.~(\ref{eq:scaling_competition})). By translating FCS into field-theoretic scaling dimensions, our results provide measurable predictions for non-equilibrium energy fluctuations in quantum simulators and phase transitions. Our formalism by combining stochastic thermodynamics and RG theory brings novel insights to the unification of classical and quantum thermodynamics under the field-theoretic framework.

\begin{acknowledgments}
This work was supported by the National Science
Foundation of China under grants 12375028 and 12521004.
\end{acknowledgments}
\bibliographystyle{apsrev4-2}
\bibliography{KZM_Work_Statistics}

\end{document}


\title{Supplemental Material: A Field-Theoretic Framework for Work Statistics and Universal Scaling in Non-equilibrium Phase Transitions}

\author{Yanbo Qiao}
\thanks{These authors contributed equally to this work.}
\affiliation{School of Physics, Peking University, Beijing 100871, China}

\author{Ruohan Xu}
\thanks{These authors contributed equally to this work.} 
\affiliation{School of Physics, Peking University, Beijing 100871, China}

\author{H. T. Quan}
\email{htquan@pku.edu.cn}
\affiliation{School of Physics, Peking University, Beijing 100871, China}
\affiliation{Collaborative Innovation Center of Quantum Matter, Beijing 100871, China}
\affiliation{Frontiers Science Center for Nano-optoelectronics, Peking University, Beijing 100871, China}

\date{\today}

\maketitle
\beginsupplement

In this supplementary material, we provide derivations supporting the main text. In Sec. I, we derive Eq. (4) in our main text. In Sec. II, we show that this generating function recovers its classical counterpart in the classical limit ($\hbar \to 0$). In Sec. III, we give numerical results for classical Ising models to verify Eq. (10) in our main text.
\section*{I. Derivation of the Generating Source}

\textit{Mass term.---}The mass term in the Keldysh action is split into the forward ($S_{m,+}$) and the backward ($S_{m,-}$) components. Substituting the piecewise protocol $r_+(t)$ and applying the time translation $t \rightarrow t+\lambda$ to the delayed interval shifts the $\lambda$-dependence onto the field component:
\begin{align}
    S_{m,+}(\lambda) &= -\frac{1}{2} \int d^d x \left[ \int_0^\lambda dt \, r_i \boldsymbol{\phi}_+^2(t) + \int_\lambda^{\tau_Q+\lambda} dt \, r(t-\lambda) \boldsymbol{\phi}_+^2(t) \right] \nonumber \\
    &= -\frac{1}{2} \int d^d x \left[ \int_0^\lambda dt \, r_i \boldsymbol{\phi}_+^2(t) + \int_0^{\tau_Q} dt \, r(t) \boldsymbol{\phi}_+^2(t+\lambda) \right].
\end{align}
The backward action $S_{m,-}(\lambda)$ with the terminal mass $r_f \equiv r(\tau_Q)$ is:
\begin{equation}
    S_{m,-}(\lambda) = \frac{1}{2} \int d^d x \left[ \int_0^{\tau_Q} dt \, r(t) \boldsymbol{\phi}_-^2(t) + \int_{\tau_Q}^{\tau_Q+\lambda} dt \, r_f \boldsymbol{\phi}_-^2(t) \right].
\end{equation}

\textit{First derivative.---}Differentiating the total mass action $S_{m}(\lambda) = S_{m,+}(\lambda) + S_{m,-}(\lambda)$ utilizes the Leibniz integral rule and the identity $\partial_\lambda \boldsymbol{\phi}_+^2(t+\lambda) = \partial_t \boldsymbol{\phi}_+^2(t+\lambda)$, we obtain the first-order source:
\begin{align}
    \frac{\partial S_m}{\partial \lambda} &= -\frac{1}{2} \int d^d x \left\{ r_i \boldsymbol{\phi}_+^2(\lambda) + \left[ r(t) \boldsymbol{\phi}_+^2(t+\lambda) \right]_0^{\tau_Q} - \int_0^{\tau_Q} dt \, \dot{r}(t) \boldsymbol{\phi}_+^2(t+\lambda) - r_f \boldsymbol{\phi}_-^2(\tau_Q+\lambda) \right\} \nonumber \\
    &= \frac{1}{2} \int d^d x \left\{ \int_0^{\tau_Q} dt \, \dot{r}(t) \boldsymbol{\phi}_+^2(t+\lambda) - r_f \left[ \boldsymbol{\phi}_+^2(\tau_Q+\lambda) - \boldsymbol{\phi}_-^2(\tau_Q+\lambda) \right] \right\},
\end{align}
 where we have applied integration by parts. Here the Keldysh contour closure imposes $\boldsymbol{\phi}_+(\tau_Q+\lambda) = \boldsymbol{\phi}_-(\tau_Q+\lambda)$, forcing the boundary term to vanish.

\textit{Total generating source.---}The first derivative reduces to the pure bulk integral. Applying further $n-1$  derivatives with respect to $\lambda$ at $\lambda=0$ and utilizing $\partial_\lambda \equiv \partial_t$ for the field component generates the $n$-th order source $S_{\text{source}}^{(n)}$ and the total generating source:
\begin{align}
    S_{\text{source}}^{(n)} \equiv \left. \frac{\partial^n S_m}{\partial \lambda^n} \right|_{\lambda=0} &= \int d^d x \int_0^{\tau_Q} dt \, \frac{1}{2} \dot{r}(t) \partial_t^{n-1} \left[ \boldsymbol{\phi}_+^2(t) \right], \\
    S_{\text{source}} &= \sum_{n=1}^\infty \frac{\lambda^n}{n!} S_{\text{source}}^{(n)}.
\end{align}
This is Eq. (4) in our main text.

\section*{II. Quantum-to-Classical Transition}

\textit{Semiclassical expansion.---}Restoring the explicit $\hbar$ dependence, the standard Keldysh rotation is defined as $\boldsymbol{\phi}_\pm = (\boldsymbol{\phi}_{\text{cl}} \pm \hbar \boldsymbol{\phi}_{\text{q}})/\sqrt{2}$. The composite field operator expands as $\boldsymbol{\phi}_+^2 = \frac{1}{2}\boldsymbol{\phi}_{\text{cl}}^2 + \mathcal{O}(\hbar)$. In the modified Keldysh contour, the temporal shift is $\lambda_t = \hbar \lambda$~\cite{funoPathIntegralApproach2018}. Therefore, the phase factor $\frac{i}{\hbar} S_\text{source}$ expands as:
\begin{equation}
    \frac{i}{\hbar} S_\text{source} = i\lambda \int d^d x \int_0^{\tau_Q} dt \, \mathcal{P}^{(1)}(t) + i\lambda \int d^d x \sum_{n=2}^{\infty} \frac{\hbar^{n-1}\lambda^{n-1}}{n!} \int_0^{\tau_Q} dt \, \mathcal{P}^{(n)}(t).
\end{equation}
Substituting $\mathcal{P}^{(1)} = \frac{1}{2}\dot{r} \boldsymbol{\phi}_+^2 = \frac{1}{4}\dot{r} \boldsymbol{\phi}_{\text{cl}}^2 + \mathcal{O}(\hbar)$ and taking the limit $\hbar \to 0$, all higher-order terms ($n \ge 2$) vanish due to the $\hbar^{n-1}$ prefactor. We identify the physical field as $\boldsymbol{\phi}_{\text{phys}} = \boldsymbol{\phi}_{\text{cl}}/\sqrt{2}$, and the phase factor reduces to the classical work~\cite{mallickFieldtheoreticApproachNonequilibrium2011, funoPathIntegralApproach2018}:
\begin{equation}
    \lim_{\hbar \to 0} \frac{i}{\hbar} S_\text{source} = i\lambda \int d^d x \int_0^{\tau_Q} dt \, \frac{1}{2}\dot{r}(t)\boldsymbol{\phi}_{\text{phys}}^2(t).
\end{equation}

\section*{III. Numerical Simulations for Classical Ising Models}
Here we simulate 2D and 3D classical Ising model (IM) with metropolis dynamics. The Hamiltonian is \cite{rotskoff2015optimal}
\begin{equation}
H\big(\{s_i\},\Lambda(t)\big)=h(t)\sum_{i=1}^ns_i-J(t)\sum_{\langle i,j\rangle}s_is_j\:,
\end{equation}
where $\langle i,j\rangle$ denotes all nearest neighbor pairs on the lattice, and the control parameter $\Lambda(t)=$ $\left(\beta h(t),\beta J(t)\right)$, varies the coupling $J$, and the external magnetic field $h$, with time. The trajectory work is defined as discrete sum over Monte Carlo timesteps
\begin{equation}
W:= \sum_{i=1}^\tau \left (\frac{\partial H}{\partial \Lambda} \right )_i \dot{\Lambda}(t)_i .
\end{equation}
Here for each step $i$ we apply a parallelized checkerboard update to acclerate calculation. The lattice size of 2D and 3D models are $L=1024$ and $L=128$, respectively. The driving protocol here is chosen to be a linear driving of $J$ from 0.5 to 1 in $\tau$ steps while keeping $T=T_c(J=1)$ and $h=0$. For each set of parameters, we run 1000 times in parallel to extract work statistics. 

\begin{figure}[htbp]
    \centering
    \begin{subfigure}[b]{0.48\textwidth}
        \centering
        \includegraphics[width=\textwidth]{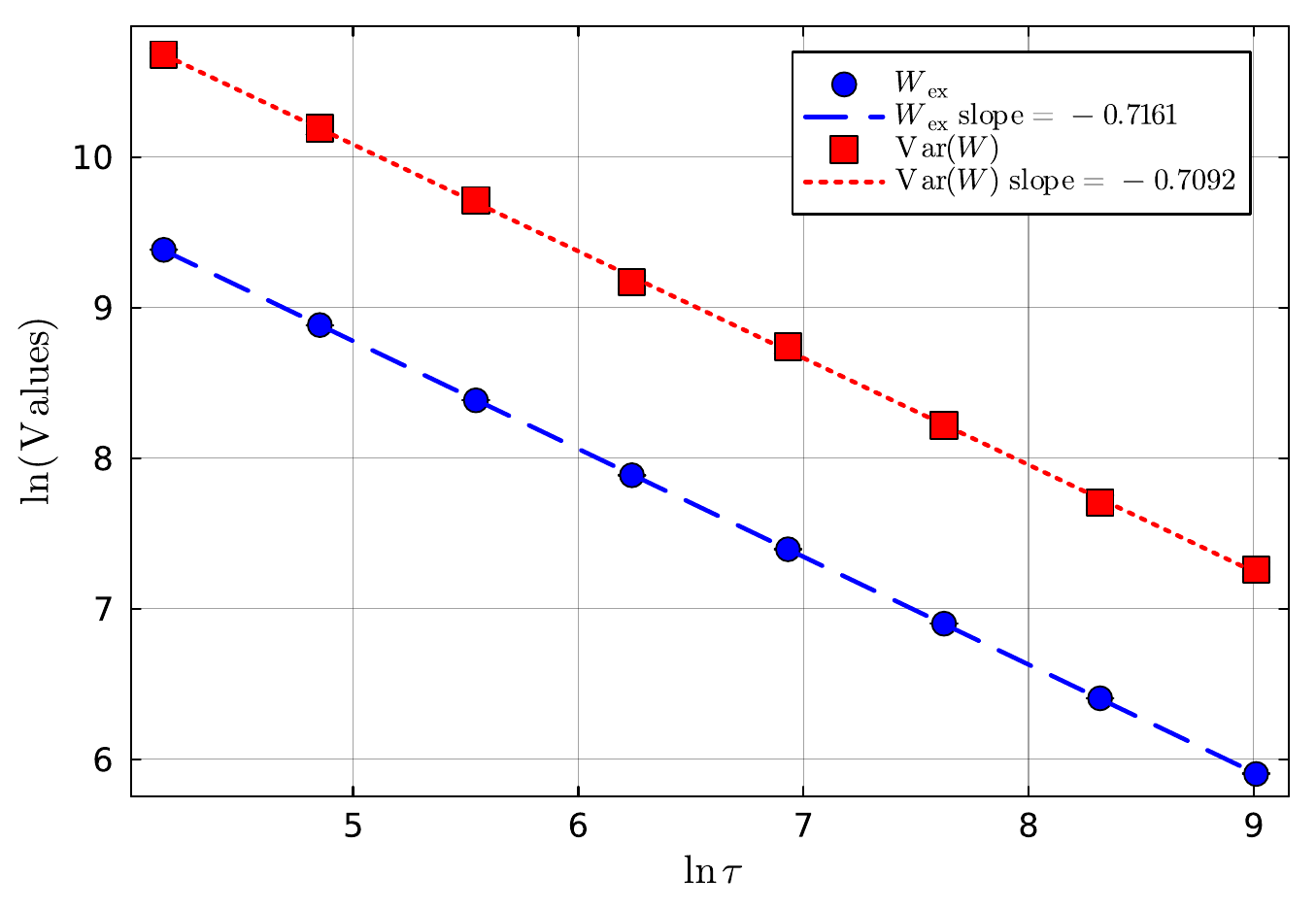}
        \caption{The excess work and variance of 2D IM}
        \label{fig:sub_left}
    \end{subfigure}
    \hfill 
    \begin{subfigure}[b]{0.48\textwidth}
        \centering
        \includegraphics[width=\textwidth]{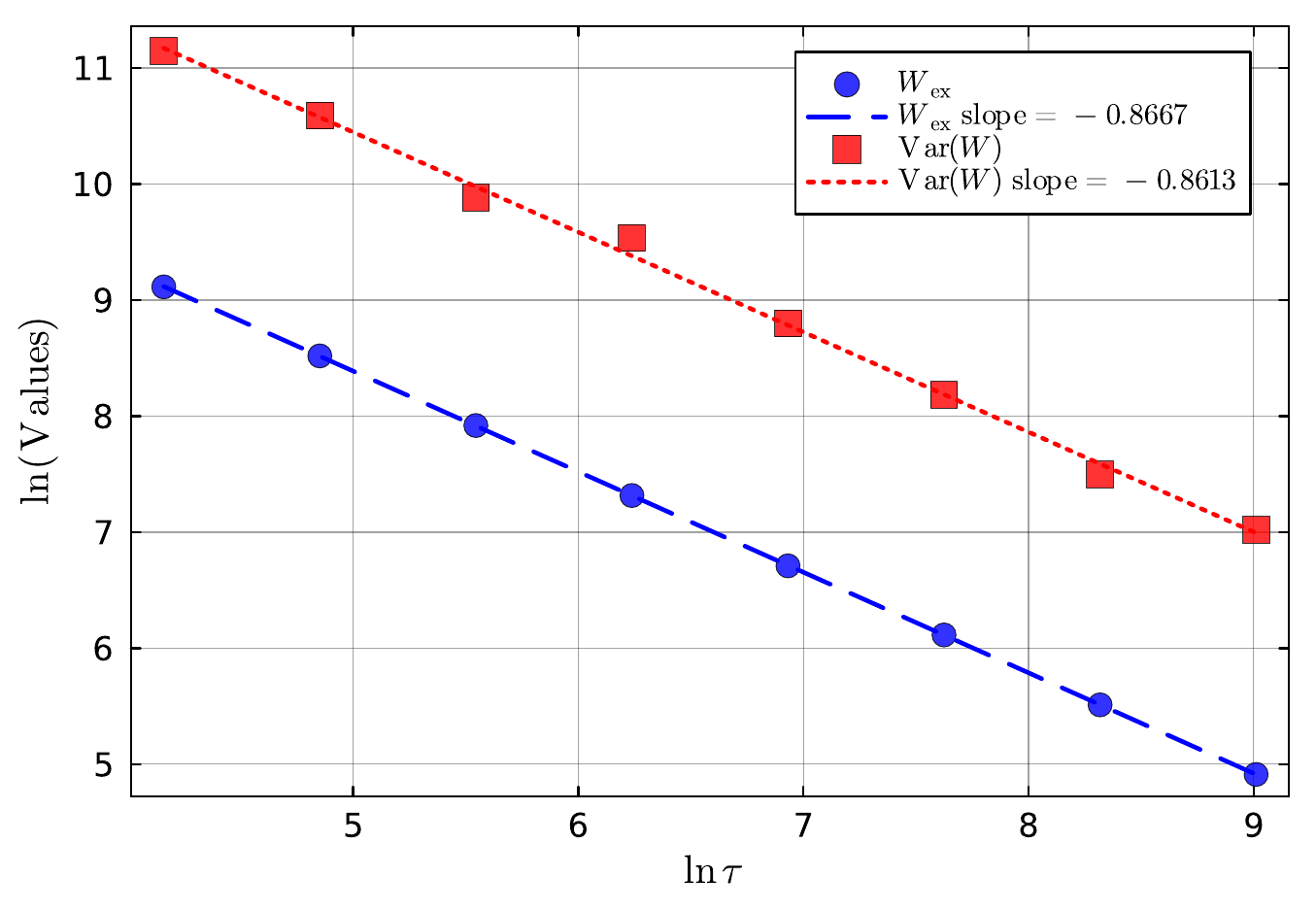}
        \caption{The excess work and variance of 3D IM}
        \label{fig:sub_right}
    \end{subfigure}
    
    \caption{Scaling of the first two cumulants for 2D (L=1024) and 3D (L=128) Ising models. The vertical axis denotes logarithm of the first and the second cumulants of excess work, and the horizontal axis denotes logarithm of the Monte Carlo timesteps.}
    \label{fig:main_total}
\end{figure}

By directly comparing with Landau-Ginzberg theory, the driving of $J$ corresponds to linear driving of $m^2$ and thus $r(t)$ in our formalism. The expected scaling exponents (from Eq. (10) in our main text) are $0.63$ and $0.83$, respectively, which are in reasonable agreement with numerical results. Here the small deviations may originate from the steeper $1/\tau$ APT background (See Eq. (11) in our main text).

\bibliographystyle{apsrev4-2}
\bibliography{KZM_Work_Statistics.bib}